# Optimal V2G and Route Scheduling of Mobile Energy Storage Devices Using a Linear Transit Model to Reduce Electricity and Transportation Energy Losses

Soon-Young Kwon, *Student Member, IEEE*, Jae-Young Park, *Student Member, IEEE*, Young-Jin Kim, *Member, IEEE*

*Abstract*—Mobile energy storage devices (MESDs) operate as medium- or large-sized batteries that can be loaded onto electric trucks and connected to charging stations to provide various ancillary services for distribution grids. This paper proposes a new strategy for MESD operation, in which their power outputs and paths are co-optimally scheduled to minimize the total energy loss in both power and transportation networks. The distances moved by MESDs and time at different locations are modeled using a set of linear equations, considering the time-varying traffic flow. The linear transit model is integrated with linearized constraints to support the reliable operation of the distribution grid. In particular, an optimal scheduling problem is formulated considering the maximum limits on incremental variations in bus voltages and line power flows for active and reactive power outputs of MESDs. A mixed-integer linear programming solver can be readily applied to the optimization problem, ensuring the global optimality of the solution. Simulation case studies are carried out under various power and transportation network conditions. The results of these case studies confirm that the proposed strategy using MESDs is effective in reducing total energy losses, compared to conventional methods using stationary batteries and plug-in electric vehicles.

## Nomenclature

The main notation used in this paper is summarized as follows.

**A. Acronyms:**
DSO    distribution system operator
MESD    mobile energy storage device
PEV    plug-in electric vehicle
RES    renewable energy source
SOC    state-of-charge
V2G    vehicle-to-grid

**B. Sets and Indices:**
$min$, $max$    subscripts for minimum and maximum values
$c$, $d$    subscripts for charging and discharging modes
$h$, $i$, $j$    indices for MESD charging stations
$s$    indices for MESD units
$a$, $b$    indices for transportation network intersections
$u$, $v$    indices for distribution network buses
$k$, $\tau$    indices for time step ($t = k \cdot t_{unit}$)

**C. Parameters:**
$t_{unit}$    unit time interval (15 minutes)
$N_I$    total number of MESD charging stations
$N_K$    number of scheduling intervals in a day ($N_K = 96$)
$N_S$    number of MESDs
$N_V$    number of distribution network buses
$E_{s,cap}$    maximum energy storage capacity of MESD unit $s$
$\Delta E_{s,max}$    maximum limit on the variation in SOC levels of MESD unit $s$ at $k = 1$ and $N_K$
$C^k$    electricity price at time step $k$
$pf_{s,min}$    minimum power factor of power output of MESD unit $s$
$\eta_s$    fuel efficiency [kWh/km] for the transit of MESD unit $s$
$\eta_{cs}$, $\eta_{ds}$    charging and discharging efficiencies of MESD unit $s$
$v_{ab}^k$    average driving speed from intersection $a$ to $b$
$d_{ab}$    distance of direct route from intersection $a$ to $b$
$p_{ab}^k$    total time taken to move from intersection $a$ to $b$
$\mathbf{S_{Ploss}}^k$, $\mathbf{S_V}^k$,    sensitivity matrices for variations in power losses, voltage
$\mathbf{S_L}^k$    magnitudes, and line power flows due to MESD power outputs at time step $k$
$\Delta\mathbf{V_{max}}$,    maximum and minimum limits on incremental bus voltage
$\Delta\mathbf{V_{min}}$    deviations
$\Delta\mathbf{L_{max}}$    maximum limits on incremental power flows
$\mathbf{A}_{ij}^k$    intersection set of the optimal path from station $i$ to $j$
$d_{ij}^k$    total distance between stations $i$ and $j$
$\gamma_{ij}^k$    total time interval required to move from station $i$ to $j$
$\mathbf{T}$    transit matrix of an MESD over a scheduling time period
$\Gamma$    maximum transit time interval $\gamma_{ij}^k$
$f_{\tau ij}^k$    binary constant reflecting the possibility of departing from station $i$ at time $k$ to arrive at station $j$ with a transit time of $\gamma_{ij}^k$
$\mathbf{F}$    coefficient matrix for the arrival state of an MESD
$P_{s,max}$, $P_{s,min}$    maximum and minimum power outputs of MESD unit $s$
$E_{s,max}$, $E_{s,min}$    maximum and minimum SOC levels of MESD unit $s$

**D. Decision Variables:**
$P_{is}^k$, $Q_{is}^k$    active and reactive power of MESD unit $s$ at station $i$
$e_{ijs}^k$    flag indicating departure of MESD unit $s$ from station $i$ to $j$

**E. State Variables:**
$e_s$    sum of $e_{ijs}^k$ values for all stations over a scheduling time period
$m_{is}^k$    flag indicating connection of MESD unit $s$ to station $i$
$y_s^k$    flag indicating moving or stationary modes of MESD $s$
$w_s^k$    flag indicating charging or discharging modes
$z_s^k$    average distance moved during time step $k$
$P_{cis}^k$, $P_{dis}^k$    MESD charging and discharging power at station $i$
$E_s^k$    SOC level of MESD unit $s$ at time step $k$
$\Delta P_{loss}^k$,    variations in network power loss, bus voltages, and line
$\Delta\mathbf{V}^k$, $\Delta\mathbf{L}^k$    power flows at time step $k$
$J$    total cost for energy loss

## I. Introduction

ENERGY storage devices (ESDs) are increasingly installed in low-voltage power grids to provide various ancillary services, including peak load shaving, reactive power support, and network investment deferral [1]–[3]. A number of studies on ESDs have been performed to assist distribution system operators (DSOs) to better utilize their ancillary services and consequently achieve more cost-effective and reliable grid operation: e.g., decreased voltage deviation and power loss and increased penetration of renewable energy sources (RESs). In particular, DSOs have adopted various strategies [1], [3] to determine the optimal locations or capacities of stationary ESDs for the most effective use of ancillary services, while avoiding excessively large costs for the installation and maintenance of ESDs. Nevertheless, it is inevitable that optimal decisions will become suboptimal as the network topology and operating conditions continue to change [4], for example due to feeder reconfiguration and RES integration.

Meanwhile, the penetration of plug-in electric vehicles (PEVs) continues to increase due to their environmental benefits. Furthermore, the development of vehicle-to-grid (V2G) technology enables the DSO to cooperate with PEV owners capable of providing distributed ancillary services that are difficult to provide using a single, stationary ESD [5]–[7]. In other words, the energy stored in PEV batteries can be used over a wide range of locations in the distribution grid in a timely manner. Optimal V2G power dispatch has been studied extensively [5]–[8], for example considering various small battery capacities, traffic congestion conditions, and departure and arrival times of individual drivers. Price- and incentive-based demand response (DR) programs [7]–[9] have also been designed to attract more PEV owners to participate in optimal V2G power dispatch, considering that owners are seldom inclined to make efforts to improve distribution grid operation. To reflect the different objectives of DSO and PEV owners, Stackelberg game theory has been adopted in DR programs [8], [9]. However, many challenges still remain with respect to the practical implementation of DR programs. For example, DSOs need to obtain battery model parameters from individual PEV owners and handle a number of auxiliary decision variables that are created during the application of Stackelberg's theory. Therefore, the operation of distribution grids often relies on centralized dispatch frameworks that use DSO-owned resources and facilities, rather than decentralized, privacy-protective schemes.

Recently, DSOs have sought to own and control mobile ESDs (MESDs) directly [10]–[14] under various operating conditions of distribution grid and transportation system. For example, electric buses and electric trucks can be shared by other DSOs via mutual assistance agreements [10], or provided by government entities with discounted prices [11]. Incentive schemes and service regulations can also be devised for DSOs to make contract with third parties and obtain investment in MESD service businesses. This is because DSOs are often instructed to ensure adequate resources for emergency and normal day-to-day operation of distribution grids [11], [12]. Considering the possibilities, several studies on MESDs have been undertaken in recent years [13], [14] where an MESD is commonly defined as a battery loaded onto an electric truck. MESDs are often characterized as having rated power and energy capacities that are larger than those of PEV batteries. Moreover, MESDs can still be connected to PEV charging stations with slight modifications to V2G chargers. The energy stored in the MESD battery is used to drive the electric truck and provide ancillary services to the distribution grid via the charging stations. Supported by local governments, DSOs have established and operated an increasing number of charging stations [5] to facilitate the use of PEVs. DSOs can also obtain temporary authority to use privately owned stations.

Various methods for MESD-based ancillary service provision were presented in [11]–[14]. For example, a dispatch framework was discussed in [11] and [12], where mobile generators were pre-positioned and dynamically allocated to minimize the expected outage duration after a natural disaster struck a distribution grid. Furthermore, a day-ahead energy management system was equipped with a two-stage optimization technique in [13] to determine the optimal V2G power outputs and positions of electric trucks, to minimize the cost of the power imported from the distribution grid. Furthermore, an analytic approach based on Markov models was adopted in [14] to evaluate the grid reliability for various penetration levels of mobile energy storage systems and intermittent energy sources, although traffic conditions were not considered.

In this paper, we propose an optimal strategy for MESD operation, where the power output and moving path schedules of the MESDs are co-optimized to minimize the total cost incurred by energy losses, based on day-ahead forecasts of load demand and traffic congestion. Energy losses consist of two parts: i.e., the incremental power loss on the distribution lines incurred by the optimal power dispatch of the MESDs, and the loss of the stored energy in the MESD batteries when driving electric trucks along the optimal paths. According to the proposed strategy, the distance moved and time schedules of the MESDs are modeled using a set of linear equations with binary flags, which are adaptively determined depending on time-varying traffic flows. The linear transit model of the MESDs is integrated with linearized constraints, ensuring stable grid operation, particularly with respect to the incremental variations in bus voltages and line power flows. The optimal scheduling problem is formulated using linear models of both transportation and power network operations, and solved using mixed-integer linear programming (MILP), which ensures the global optimality of the schedule. Simulation case studies are carried out using small- and large-scale test beds under various conditions, which are determined based on factors such as traffic time delays, MESD power and energy capacities, and incremental changes in the bus voltage and line power flow. The results of the case studies have demonstrated that the proposed strategy assists the DSO more effectively in reducing the total energy loss cost, compared to the conventional strategies using stationary ESDs and PEVs.

The contributions of this paper are summarized as follows:
• The moving distance and the transit time period of an MESD are modeled using a set of linear equations with binary flags that indicate the connections of MESDs to charging stations. Given time- and location-varying traffic congestion data, the optimal path of each MESD is determined by calculating the fastest routes between intersections. The transit model is expanded in the form of linear matrix inequality constraints; this allows scheduling of optimal paths for individual MESDs over a period of many hours.
• Using the linearized transit model, an optimization problem is formulated to simultaneously schedule both the optimal

power outputs and moving paths of MESDs, facilitating analysis of mutual interactions between the power and traffic networks. To the best of our knowledge, this is the first study to develop and apply a linearized transit model to an optimal scheduling problem, so that it can be readily solved via MILP while ensuring global optimality of the solution with computational time efficiency.

• Comparative case studies are carried out under various conditions of power and transportation network operations. The case study results demonstrate that the proposed strategy is more effective in reducing the cost for total energy loss, compared to the conventional strategies employing stationary ESDs and PEVs.

Section II defines the linear transit model of the MESDs. Section III then presents the optimization problem using the transit model. Section IV discusses the case studies and results for the proposed and conventional strategies. Section V provides discussions, and Section IV concludes the paper.

## II. LINEARIZED TRANSIT MODEL OF MESDs

During the scheduling time period of $1 \leq k \leq N_K$, the MESDs are connected to battery charging stations and move as required between stations along traffic routes. The charging stations that MESDs depart from and arrive at are located at intersections on the road map and connected to buses in the distribution network. In other words, the stations are candidate sites where MESDs can be interfaced with the distribution grid via battery charging and discharging apparatus. For simplicity, it is assumed that the stations are already equipped with interface infrastructure, such as AC/DC bi-directional power converters [15].

For the route distance $d_{ab}$ between intersections $(a, b)$, the average speed $v_{ab}^k$ at time $k$ can be predicted using historical data that have been collected from speed sensors installed on each route [16]. Note that $v_{ab}^k$ reflects traffic congestion at the time step $k$, whereas $d_{ab}$ is fixed for any transportation network. Using the forecast $v_{ab}^k$, the average time period $p_{ab}^k$ for transit of an MESD at time step $k$ between intersections $(a, b)$ can be calculated in advance as:

$$p_{ab}^k = d_{ab} / v_{ab}^k, \quad \forall k. \quad (1)$$

The intersection set $\mathbf{A}_{ij}^k$ for the optimal, fastest path from station $i$ to station $j$ at time $k$ can then be pre-determined using Dijkstra's algorithm [12], based on the sum of the minimum value of $p_{ab}^k$ (i.e., the shortest transit time) for each pair of intersections $(a, b)$ on the path from station $i$ to station $j$. For example, if the optimal path connecting stations $i$ and $j$ consists of routes between intersections $(a_\sigma, a_{\sigma+1})$ for $1 \leq \sigma \leq 3$, then $\mathbf{A}_{ij}^k$ is specified as a combination of partial routes: i.e., $\mathbf{A}_{ij}^k = \{(i (= a_1), a_2), (a_2, a_3), (a_3, j (= a_4))\}$. Note that, due to the time-varying traffic congestion, the optimal path from station $i$ to station $j$ at time $k$ can vary during the scheduling time period. The optimal distance set $\mathbf{D}^k$, corresponding to $\mathbf{A}_{ij}^k$ for all stations $1 \leq i \leq N_I$, can then be expressed as:

$$\mathbf{D}^k = \underbrace{\begin{bmatrix} d_{11}^k & \cdots & d_{1N_I}^k \\ \vdots & d_{ij}^k & \vdots \\ d_{N_I 1}^k & \cdots & d_{N_I N_I}^k \end{bmatrix}}_{N_I \times N_I}, \quad 1 \leq i, j \leq N_I, \forall k, \quad (2)$$

where $d_{ij}^k$ is the distance that takes the shortest time to transit from station $i$ to station $j$ at time step $k$ for the optimal intersection set $\mathbf{A}_{ij}^k$. This implies that all diagonal elements $d_{ii}^k$ in $\mathbf{D}^k$ are zero for all $k$ (i.e., $d_{ii}^k = 0$ for $1 \leq i \leq N_I$ and $1 \leq k \leq N_K$).

Furthermore, the normalized time interval $\gamma_{ij}^k$ required for the transit from station $i$ to station $j$ at time $k$ is calculated using the unit time interval $t_{unit}$ (i.e., 15 minutes) as:

$$\gamma_{ij}^k = \begin{cases} \left\lceil \dfrac{1}{t_{unit}} \sum_{(a,b) \in \mathbf{A}_{ij}^k} p_{ab}^k \right\rceil, & \forall i \neq j, \forall k, \\ 0, & \forall i = j, \forall k, \end{cases} \quad (3)$$

where $\lceil \cdot \rceil$ represents the smallest integer greater than or equal to $\cdot$. In (3), $\gamma_{ij}^k$ is an integer constant that can be estimated in advance using forecasts of $v_{ab}^k$ for intersections $(a, b) \in \mathbf{A}_{ij}^k$. For $i = j$, $\gamma_{ij}^k$ is zero, as for $d_{ij}^k$ of (2). The $\lceil \cdot \rceil$ is introduced in (3) to allow the DSO to make decisions on MESD operation at each time step $k$. Specifically, $p_{ab}^k$, and hence the total time interval $\sum_{(a, b)} p_{ab}^k$ required to move from station $i$ to station $j$ are likely, in practice, to have continuous real values. The DSO determines the optimal power outputs and transit paths of MESDs at integer time steps $k = t/t_{unit} = 1, \cdots, N_K \, (= 96)$. As shown in Fig. 1, when an MESD arrives at station $j$ at time $t_{MESD}$, the DSO schedules optimal operation of the MESD at a time $t_{DSO} \geq t_{MESD}$. Using the operational character $\lceil \cdot \rceil$, the time $t_{MESD}$ at which the MESD changes its traveling state can be synchronized with the time $t_{DSO}$ at which the DSO makes the optimal decision (i.e., $t_{DSO}/t_{unit} = \lceil t_{MESD}/t_{unit} \rceil$). In practice, the MESD is expected to be in the idle mode during the time period $t_{DSO} - t_{MESD}$.

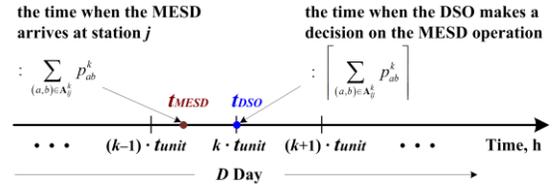

Fig. 1. Comparison of the actual time interval $\sum_{(a,b)} p_{ab}^k$ required for an MESD to transit from station $i$ to $j$ with the discretized time interval $\left\lceil \sum_{(a,b)} p_{ab}^k \right\rceil$ used for optimal decision-making by the DSO at each time step $k$.

In addition, the MESD unit $s$ departing towards station $j$ from station $i$ at time $k$ is not connected to any station during the transit time period, from $k$ to $k + \gamma_{ij}^k$, because $\gamma_{ij}^k$ in (3) represents the minimum time interval that it takes to arrive at station $j$. This can be expressed equivalently as:

$$m_{is}^k + \frac{1}{N_I \cdot \Gamma} \sum_{1 \leq \lambda_{ij}^k \leq \gamma_{ij}^k} \sum_{j \neq i} m_{js}^{(k+\lambda_{ij}^k)} \leq 1, \quad \forall i, \forall j, \forall s, \quad (4)$$

$$\Gamma = \max_{i,j,k} \{\gamma_{ij}^k\}, \quad \forall i, \forall j, \forall k, \quad (5)$$

where $m_{is}^k$ represents the connection state of MESD unit $s$ at station $i$ at time $k$. If the MESD is connected, $m_{is}^k$ is set to one; otherwise, it is equal to zero. In (5), $\Gamma$ is defined as the maximum value of the transit time interval $\gamma_{ij}^k$ for all cases of $i$, $j$, and $k$. It can be seen in (4) and (5) that when $m_{is}^k = 0$ for departure station $i$, $m_{js}^k$ for any arrival station $j \neq i$ can be set to either 1 or 0 during the transit time period, because $N_I \cdot \Gamma$ is greater than the sum of $m_{js}^k$ for all $j$ and $\lambda_{ij}^k$. In contrast, when $m_{is}^k = 1$, $m_{js}^k$ can only be zero during the transit time; $1 \leq \lambda_{ij}^k \leq \gamma_{ij}^k$.

Based on (4) and (5), the transit matrix $\mathbf{T}$ is then defined to develop the transit model of MESD unit $s$ during the scheduling time period $1 \leq k \leq N_K$ as:

$$\mathbf{T} = \underbrace{\begin{bmatrix} \mathbf{I} & \mathbf{F}_1^1 & \mathbf{F}_2^1 & \cdots & \mathbf{F}_\Gamma^1 & \mathbf{O} & \mathbf{O} & \cdots & \mathbf{O} \\ \mathbf{O} & \mathbf{I} & \mathbf{F}_1^2 & \mathbf{F}_2^2 & \cdots & \mathbf{F}_\Gamma^2 & \mathbf{O} & \cdots & \mathbf{O} \\ \vdots & \vdots & \vdots & \vdots & \vdots & \vdots & \vdots & & \vdots \\ \mathbf{O} & \mathbf{O} & \mathbf{O} & \mathbf{O} & \mathbf{O} & \cdots & \mathbf{O} & \mathbf{I} & \mathbf{F}_1^{N_K-1} \end{bmatrix}}_{N_I(N_K-1) \times N_I N_K}, \quad (6)$$

where 
$$\mathbf{F}_\tau^k = \frac{1}{N_I \cdot \Gamma} \underbrace{\begin{bmatrix} 0 & f_{\tau 12}^k & \cdots & f_{\tau 1 N_I}^k \\ f_{\tau 21}^k & 0 & \cdots & f_{\tau 2 N_I}^k \\ \vdots & \vdots & f_{\tau ij}^k & \vdots \\ f_{\tau N_I 1}^k & f_{\tau N_I 2}^k & \cdots & 0 \end{bmatrix}}_{N_I \times N_I}, \quad (7)$$

$$\forall \tau \in \{1, 2, \ldots, \gamma_{ij}^k, \ldots, \Gamma\},$$

$$f_{\tau ij}^k \in \{0, 1\}, \quad \forall i, \forall j, \forall k, 1 \leq \tau \leq \Gamma. \quad (8)$$

In (6), $\mathbf{O}$ and $\mathbf{I}$ are the zero and identity matrices, respectively, with dimensions of $N_I \times N_I$. In (7), $\mathbf{F}_\tau^k$ is a binary matrix of dimensions $N_I \times N_I$, divided by the scalar factor $N_I \cdot \Gamma$. Consequently, the transit matrix $\mathbf{T}$ is of dimensions $N_I \cdot (N_K-1) \times N_I \cdot N_K$. The sub-matrix $\mathbf{I}$ corresponds to the coefficients in the first term of (4) for the MESD departure state from station $i$, and $\mathbf{F}_\tau^k$ accounts for the coefficients in the second term of (4) for the arrival state at station $j$. In other words, the elements of $\mathbf{T}$ are the coefficients of $m_{is}^k$ and $m_{js}^k$ in (4). Specifically, in (7), $f_{\tau ij}^k$ is a binary constant indicating the probability that an MESD departs from station $i$ at time $k$ to arrive at station $j$, which requires a transit time of $\gamma_{ij}^k$. Given $\gamma_{ij}^k$, $f_{\tau ij}^k$ is set to 1 for $1 \leq \tau \leq \gamma_{ij}^k$, indicating that the MESD can arrive at station $j$ at time $k + \gamma_{ij}^k + 1$; otherwise, the value is zero. Note that $f_{\tau ij}^k$ and hence $\mathbf{T}$ are pre-determined constants derived using the forecast $v_{ab}^k$ and the pre-estimation $\gamma_{ij}^k$; these parameters reflect time-varying traffic congestion over a scheduling time period.

Using (6)–(8), the transit model (4) and (5) can then be expanded to (9) and (10) for the scheduling time period as:

$$\mathbf{T} \cdot \mathbf{M}_s \leq \mathbf{U} = \underbrace{[1, \cdots, 1]^T}_{N_I(N_K-1)}, \quad \forall s, \quad (9)$$

where $\mathbf{M}_s = \underbrace{\left[ m_{1s}^1, m_{2s}^1, \cdots, m_{N_I s}^1, \cdots, m_{1s}^{N_K}, m_{1s}^{N_K}, \cdots, m_{N_I s}^{N_K} \right]^T}_{N_I \cdot N_K}. \quad (10)$

Note that (9) is in the form of a linear matrix inequality constraint. The constant matrix $\mathbf{T}$ in (6) is multiplied by the decision variable vector $\mathbf{M}_s$, which consists of connection flags $m_{is}^k$ for all stations $1 \leq i \leq N_I$ over the scheduling time period $1 \leq k \leq N_K$. The column vector $\mathbf{M}_s$ has dimensions of $N_I \cdot N_K$, as shown in (10). For all MESD units, (6)–(10) remain the same: i.e., $\mathbf{T}_{s=1} = \cdots = \mathbf{T}_{s=N_S} = \mathbf{T}$. In other words, the transit of an MESD does not affect the traffic congestion on its route and, consequently, the required transit time intervals of other MESDs: i.e., the transits of the MESDs are independent. The transit model for all MESD units can then be expanded as:

$$\underbrace{\begin{bmatrix} \mathbf{T} & \mathbf{O} & \mathbf{O} \\ \mathbf{O} & \ddots & \mathbf{O} \\ \mathbf{O} & \mathbf{O} & \mathbf{T} \end{bmatrix}}_{N_I(N_K-1)N_S \times N_I N_K N_S} \cdot \underbrace{\begin{bmatrix} \mathbf{M}_{s=1} \\ \vdots \\ \mathbf{M}_{s=N_S} \end{bmatrix}}_{N_I N_K N_S \times 1} \leq \underbrace{\begin{bmatrix} \mathbf{U} \\ \vdots \\ \mathbf{U} \end{bmatrix}}_{N_I(N_K-1)N_S \times 1}, \quad (11)$$

where the left-hand side term $\mathbf{T} \cdot \mathbf{M}_s$ of (9) for each $s = 1, 2, \cdots, N_S$ are integrated, forming a large diagonal matrix with sub-matrices $\mathbf{T}$ and $\mathbf{O}$. The total dimensions are $N_I \cdot (N_K - 1) \cdot N_S \times N_I \cdot N_K \cdot N_S$, as shown in (11); given the independence of the MESD transits, each off-diagonal sub-matrix is set to $\mathbf{O}$ of dimensions $N_I \cdot (N_K - 1) \times N_I \cdot N_K$. Similarly, $\mathbf{M}_s$ and $\mathbf{U}$ are stacked for $1 \leq s \leq N_S$, yielding large column vectors with dimensions of $N_I \cdot N_K \cdot N_S$ and $N_I \cdot (N_K - 1) \cdot N_S$, respectively.

To complete the linear transit model (11), the binary variable $e_{ijs}^k \in \{0, 1\}$ is defined as the departure flag, which assists the DSO in estimating the distance moved by the MESDs and hence the energy lost by driving the MESDs along the optimal paths. Specifically, $e_{ijs}^k$ is set to one only when the MESD unit $s$ departs from station $i$ to station $j$ at time $k$; otherwise, it is equal to zero. Therefore, the possible number of journeys for each MESD at time $k$ is limited to one, as shown in (12). Consequently, (13) shows the total number of MESD transits during the scheduling period.

$$\sum_{i}^{N_I} \sum_{j \neq i}^{N_I} e_{ijs}^k \leq 1, \quad \forall k, \forall s \quad (12)$$

$$e_s = \sum_{k}^{N_K} \sum_{i}^{N_I} \sum_{j \neq i}^{N_I} e_{ijs}^k, \quad \forall s \quad (13)$$

Furthermore, $e_{ijs}^k$ can be related to $m_{is}^k$ as follows:

$$e_{ijs}^k + \sum_{h=1, h \neq i, j}^{N_I} e_{ihs}^k \geq \left( m_{is}^k - m_{is}^{k+1} \right) - \left( m_{js}^{k+\gamma_{ij}^k} - m_{js}^{k+\gamma_{ij}^k+1} \right) - 1, \quad (14)$$
$$\forall i \neq j, \forall s, \forall k,$$

$$e_{ijs}^k \leq \left( m_{is}^k + m_{js}^{k+\gamma_{ij}^k+1} \right) \big/ 2, \quad \forall i, \forall j, \forall s, \forall k, \quad (15\text{-a})$$

$$e_{ihs}^k \leq \left( m_{is}^k + m_{hs}^{k+\gamma_{ih}^k+1} \right) \big/ 2, \quad \forall i, \forall h, \forall s, \forall k, \quad (15\text{-b})$$

$$y_s^k = 1 - \sum_{i=1}^{N_I} m_{is}^k, \quad \forall k, \forall s, \quad (16)$$

$$y_s^k = \sum_{j=1}^{N_I} \sum_{i=1, i \neq j}^{N_I} \sum_{\tau=1}^{k-1} f_{(k-\tau)ij}^\tau e_{ijs}^\tau, \quad \forall k \geq 2, \forall s, \quad (17)$$

$$z_s^k = \sum_{j=1}^{N_I} \sum_{i \neq j}^{N_I} \sum_{\tau=1}^{k-1} d_{ij}^\tau \left( \gamma_{ij}^\tau \right)^{-1} f_{(k-\tau)ij}^\tau e_{ijs}^\tau, \quad \forall k \geq 2, \forall s. \quad (18)$$

Specifically, in (14), the term on the right-hand side accounts for the connection state at station $i$ at time $k$, and at station $j$ at time $k + \gamma_{ij}^k$ for the transit of MESD unit $s$. The MESD transit is achieved using either direct (i.e., from the station $i$ to $j$) or indirect (i.e., from the station $i$ to $h$, and to $j$) movement. The left-hand term of (14) takes the two types of movement into consideration separately, whereas the right-hand term remains equal to one for both types of movement. In other words, only the connection states at the departure and arrival times are considered in the right term. For example, departure of an MESD at time $k$ is equivalent to the condition that the MESD is connected to a charging station at time $k$ and disconnected at time $k + 1$ (i.e., $m_{is}^k = 1$ and $m_{is}^{k+1} = 0$). Similarly, arrival at time step $k + \gamma_{ij}^k + 1$ is equivalent to the condition that the MESD is disconnected at time $k + \gamma_{ij}^k$ and connected at time $k + \gamma_{ij}^k + 1$ (i.e., $m_{js}^{k+\gamma_{ij}^k} = 0$ and $m_{js}^{k+\gamma_{ij}^k+1} = 1$).

In the case of direct movement, the right terms in (15-a) and (15-b) become equal to 1 and 0.5, respectively; this results in $e_{ijs}^k = 1$ and $e_{ihs}^k = 0$ based on (14), clearly indicating direct movement from station $i$ to $j$. Note that (12) is also satisfied. In the case of indirect movement, both right-hand terms are equal to one, which requires that we further examine (16) and (17) to determine $e_{ijs}^k$ and $e_{ihs}^k$. Specifically, $y_s^k$ is defined as the moving flag to indicate whether or not MESD unit $s$ is on the move at time $k$. As shown in (16), $y_s^k$ is set to one when the MESD is not connected to any station at time $k$; otherwise, $y_s^k$ is equal to zero. Constraint (17) also specifies the condition on $y_s^k$; it is only set to one in the case such that the MESD departed from any station earlier (i.e., $\sum_i e_{ijs}^\tau = 1$ for $\tau < k$) and has not yet arrived at any other station (i.e., $\sum_j f_{(k-\tau)ijs}^\tau = 1$). In the case of indirect movement, enforcing $e_{ijs}^k$ to be equal to one results in a contradiction between (16) and (17); i.e., $y_s^k = 0$ in (16) and 1 in (17) at time $k + \gamma_{ih}^k + 1$. This leaves one option where $e_{ijs}^k = 0$ and hence $e_{ihs}^k = 1$ in (14), correctly indicating the indirect movement from station $i$ to $h$ and then to $j$. Moreover, using (17), (18) represents the travel distance of MESD unit $s$ during a unit time interval in the average sense [i.e., $d_{ij}^\tau \cdot (\gamma_{ij}^\tau)^{-1}$] when MESD unit $s$ is on a direct or indirect route at time $k$; under both conditions, $f_{(k-\tau)ijs}^\tau \cdot e_{ijs}^\tau$ is 1. Fig. 2 shows an illustrative example of the relationship between the binary flags (i.e., $m_{is}^k$, $e_{ijs}^k$, and $y_s^k$) of MESD unit $s$ during the transit period of $\gamma_{ij}^k$ under stationary, direct movement, and indirect movement conditions. Fig. 2 also shows the average distance $z_s^k$ traveled for a unit time interval under each moving condition.

It should be noted that using the forecast of $v_{ab}^k$, the DSO can pre-determine the values of the constant parameters or matrices $p_{ab}^k$, $\mathbf{A}_{ij}^k$, $d_{ij}^k$, $\mathbf{D}^k$, $\gamma_{ij}^k$, $\Gamma$, $f_{\tau ij}^k$, $\mathbf{F_\tau^k}$, and $\mathbf{T}$ in (1)–(11) one day in advance (i.e., before solving the optimization problem, as discussed in Section III). Therefore, (12)–(18) are linear constraints with the variables $m_{is}^k$, $e_{ijs}^k$, $y_s^k$, and $z_s^k$ to be determined, and consequently can be readily integrated with the linear transit model (11) into the optimization problem.

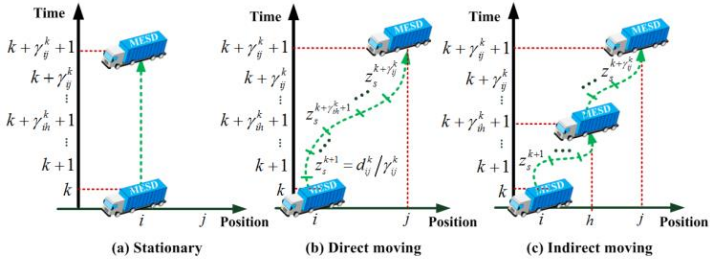

Fig. 2. Illustrative example of the relationship between $m_{is}^k$, $e_{ijs}^k$, and $y_s^k$ under the stationary and moving conditions of the MESD.

## III. OPTIMIZATION PROBLEM FORMULATION

### A. Proposed strategy for optimal MESD operation

Fig. 3 shows an overall schematic diagram of the proposed strategy for optimally scheduling the MESD power outputs and transit paths over a 24-hour period. For optimal scheduling, it is assumed that the DSO can forecast the net load demand and has access to the information on traffic time delays for the day ahead [13], [17]. The co-optimization problem is formulated by integrating the linear transit model of the MESDs (developed in Section II) with the linearized constraints on the variations in the MESD power outputs and stored energy levels, and on the corresponding effects on the distribution grid operation (discussed in Section III-B). An off-the-shelf MILP solver can be readily applied to the optimization problem, ensuring the global optimality of the solution. Note that the proposed strategy can be implemented as a single-level decision model, rather than binary-level models, because the DSO controls the MESDs directly, unlike the PEVs, while also taking full responsibility for stable operation of the distribution grid.

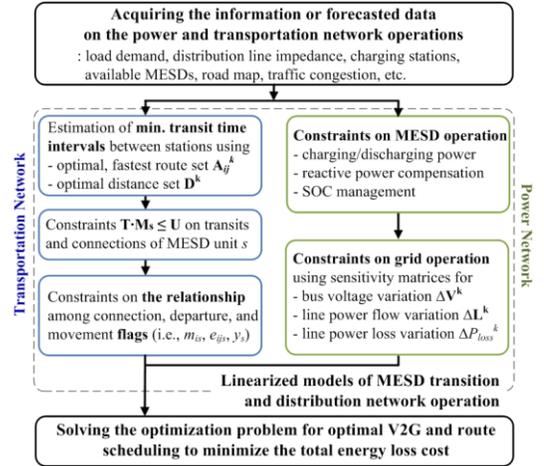

Fig. 3. Schematic diagram of the proposed power dispatch and route planning.

### B. Objective function and linearized constraints

The optimal scheduling of the MESD power outputs and transit paths can be achieved by solving:

$$\underset{P_{is}^k, Q_{is}^k, e_{ijs}^k}{\arg\min} \quad J = \sum_{k=1}^{N_K} C^k \Delta P_{loss}^k t_{unit} + \sum_{k=1}^{N_K} \sum_{j=1}^{N_I} \sum_{i=1, i\neq j}^{N_I} \sum_{s=1}^{N_S} \eta_s C^k d_{ij}^k e_{ijs}^k, \quad (19)$$

subject to

$$m_{is}^k \cdot P_{s,\min} \leq P_{is}^k \leq m_{is}^k \cdot P_{s,\max}, \quad \forall k, \forall i, \forall s, \quad (20)$$

$$P_{is}^k = P_{cis}^k + P_{dis}^k, \quad \forall k, \forall i, \forall s, \quad (21)$$

$$(1 - w_s^k) \cdot P_{s,\min} \leq P_{cis}^k \leq 0, \quad 0 \leq P_{dis}^k \leq w_s^k \cdot P_{s,\max}, \quad \forall k, \forall i, \forall s, \quad (22)$$

$$-\sqrt{1 - pf_{s,\min}^2} \cdot (P_{dis}^k - P_{cis}^k) \leq pf_{s,\min} \cdot Q_{is}^k$$
$$\leq \sqrt{1 - pf_{s,\min}^2} \cdot (P_{dis}^k - P_{cis}^k), \quad \forall k, \forall i, \forall s, \quad (23)$$

$$E_s^{k-1} - E_s^k = \frac{(\eta_{cs} P_{cis}^k + \eta_{ds} P_{dis}^k) t_{unit}}{E_{s,cap}} + \frac{\eta_s z_s^k}{E_{s,cap}}, \quad \forall k, \forall s, \quad (24)$$

$$E_{s,\min} \leq E_s^k \leq E_{s,\max}, \quad \forall k, \forall s, \quad (25)$$

$$-\Delta E_{s,\max} \leq E_s^{N_K} - E_s^0 \leq \Delta E_{s,\max}, \quad \forall s, \quad (26)$$

$$\Delta P_{loss}^k = \begin{bmatrix} \mathbf{S_{Ploss,P}^k} & \mathbf{S_{Ploss,Q}^k} \end{bmatrix} \begin{bmatrix} \mathbf{P^k} \\ \mathbf{Q^k} \end{bmatrix}, \quad \forall k, \quad (28)$$

$$\Delta \mathbf{V_{min}^k} \leq \Delta \mathbf{V^k} = \begin{bmatrix} \mathbf{S_{V,P}^k} & \mathbf{S_{V,Q}^k} \end{bmatrix} \begin{bmatrix} \mathbf{P^k} \\ \mathbf{Q^k} \end{bmatrix} \leq \Delta \mathbf{V_{max}^k}, \quad \forall k, \quad (29)$$

$$\Delta \mathbf{L_{min}^k} \leq \Delta \mathbf{L^k} = \begin{bmatrix} \mathbf{S_{L,P}^k} & \mathbf{S_{L,Q}^k} \end{bmatrix} \begin{bmatrix} \mathbf{P^k} \\ \mathbf{Q^k} \end{bmatrix} \leq \Delta \mathbf{L_{max}^k}, \quad \forall k, \quad (30)$$

and (1)–(18), (27).

In (19), the objective function is defined as the sum of two terms; the first represents the incremental energy loss in distribution lines attributable to the power outputs of MESDs, and the second is the loss of energy stored in MESD batteries because of movement of the electric trucks. These are then multiplied by the time-varying electricity price $C^k$ to calculate the total energy loss cost $J$. The objective function (19) is linear, because $d_{ij}^k$ is a constant pre-determined based on the $v_{ab}^k$ forecast, as discussed in Section II. This corresponds to the task shown in Fig. 3; it is necessary to estimate the intersection set $\mathbf{A}_{ij}^k$, the distance set $\mathbf{D}^k$, and the minimum time intervals $\gamma_{ij}^k$ for transit between stations. Note that $C^k$ and $t_{unit}$ are also constants. The DSO aims to minimize $J$ by calculating optimal profiles of the active and reactive power outputs $P_{is}^k$ and $Q_{is}^k$, and the movements $e_{ijs}^k$ between charging stations for all MESD units during the scheduling time period $1 \leq k \leq N_K$.

Constraint (20) specifies that, for MESD unit $s$ linked to station $i$, $P_{is}^k$ is restricted to within the rated charging and discharging power capacities: i.e., $P_{s,min} < 0$ and $P_{s,max} > 0$, respectively. If the MESD is not connected (i.e., $m_{is}^k = 0$), $P_{is}^k$ is maintained at zero. Moreover, in (21), $P_{is}^k$ is represented as the sum of the charging power $P_{cis}^k < 0$ and discharging power $P_{dis}^k > 0$. For the V2G service provision, each MESD can operate adaptively, in either charging or discharging mode, to account for variations in the load demand and RES power generation. The operating mode of MESD unit $s$ is determined using (22) with the binary variable $w_s^k \in \{0, 1\}$. Specifically, if $w_s^k$ is set to one, the MESD operates in the discharging mode. In other words, $P_{cis}^k$ is constant at zero and $P_{is}^k$ then becomes equal to $P_{dis}^k$, ranging from zero to $P_{s,max}$ (i.e., for $w_s^k = 1$, $P_{cis}^k = 0$ and $0 \leq P_{is}^k = P_{dis}^k \leq P_{s,max}$). Similarly, for $w_s^k = 0$, the MESD operates in the charging mode (i.e., $P_{dis}^k = 0$ and $P_{s,min} \leq P_{is}^k = P_{cis}^k \leq 0$). Note that (22) represents a set of mixed-integer linear inequality constraints, because in general, $P_{s,min}$ and $P_{s,max}$ are constants in specifications for real batteries. Furthermore, the MESDs should comply with regulations on the minimum power factor $pf_{s,min}$, which is consistent with the common requirement imposed on RESs and diesel generators in distribution networks [5]. For $pf_{s,min}$, the dispatch range of $Q_{is}^k$ is specified in (23) for the discharging and charging modes, where $(P_{dis}^k - P_{cis}^k)$ becomes equal to $P_{dis}^k$ and $-P_{cis}^k$, respectively; note that both terms are positive. In this study, $pf_{s,min}$ is set to a constant 0.95, ensuring the linear inequality of (23).

Furthermore, (24)–(26) represent constraints on the state-of-charge (SOC) levels. Specifically, (24) includes two terms for estimating the SOC variation per unit time interval. The first term deals with the effects of the charging and discharging power on SOC variation, and the second term represents the decrease in the SOC level for a driving distance $z_s^k$, calculated using (18). In (24), we consider that the battery charging and discharging efficiencies [13] and the fuel efficiency [18] (i.e., $\eta_{cis}$, $\eta_{dis}$, and $\eta_s$, respectively) are all constant, yielding a linear equality constraint (24). Moreover, the maximum energy storage capacity $E_{s,cap}$ is often constant in specifications for real-world batteries. Using the estimate in (24), (25) specifies the maximum and minimum limits on $E_s^k$ to prevent the battery over-charging and over-discharging, respectively. Constraint (26) also requires the difference between the SOC levels at $k = 0$ and $N_K$ to remain within the acceptable range for continuous MESD operation over the next day.

Furthermore, the column vector $[\mathbf{P}^k, \mathbf{Q}^k]^T$ is established in (27) using $P_{is}^k$ and $Q_{is}^k$ as elements, based on the locations of charging stations in the distribution grid. As an example, (27) shows $[\mathbf{P}^k, \mathbf{Q}^k]^T$ for the test grid, discussed in Section IV-A, where five stations are connected to Buses 6, 11, 19, 28, and 31, respectively. The vector contains $2 \cdot N_V$ elements in the case of a 3-phase balanced, $N_V$-bus grid. Each set of two elements represents $[\sum_s P_{is}^k, \sum_s Q_{is}^k]^T$ for the corresponding bus to which station $i$ is connected. Using $[\mathbf{P}^k, \mathbf{Q}^k]^T$, the DSO can readily estimate the incremental variations in the network power loss $\Delta P_{loss}^k$, bus voltage magnitudes $\Delta \mathbf{V}^k$, and line power flows $\Delta \mathbf{L}^k$ for the active and reactive power outputs of the MESDs, assuming that the power outputs are sufficiently smaller than the net load demand. As shown in (28)–(30), $\Delta P_{loss}^k$, $\Delta \mathbf{V}^k$, and $\Delta \mathbf{L}^k$ are calculated using the sensitivity matrices $\mathbf{S}_{Ploss}^k$, $\mathbf{S}_V^k$, and $\mathbf{S}_L^k$, respectively, which can be derived from general power flow equations [19]. The constraints (29) and (30) require $\Delta \mathbf{V}^k$ and $\Delta \mathbf{L}^k$ to be maintained between the maximum and minimum limits for all buses and lines, respectively.

For the proposed strategy, the optimization problem is completed by integrating (1)–(18) into a set of constraints. The optimization problem can be applied to various traffic and power networks, including different numbers and type of MESDs, with the constants being modified accordingly: for example, $\mathbf{A}_{ij}^k$, $\mathbf{D}^k$, $\gamma_{ij}^k$, $f_{\tau ij}^k$, $\mathbf{T}$, $P_{s,max}$, $P_{s,min}$, $\eta$, and $E_{s,cap}$ for the MESDs of the transportation network and $C^k$, $\mathbf{S}_{Ploss}^k$, $\mathbf{S}_V^k$, and $\mathbf{S}_L^k$ for the distribution grid. Because the DSO is responsible for stable and cost-effective operation of MESDs and distribution grids within its territory, the DSO is expected to be aware of the constants in advance, ensuring linearity of the objective function and the constraints. This is consistent with current practices for the operation of grid-connected batteries.

### IV. CASE STUDIES AND RESULTS

#### A. Test bed and simulation conditions

The proposed strategy was tested using the IEEE 34-Node Test Feeder with slight modifications based on [13] and [15], as shown in Fig. 4. The test grid includes wind turbines (WTs) at Buses 10 and 12 and photovoltaic (PV) arrays at Bus 20. Fig. 5(a) and (b) show the forecast profiles of the load demand and RES power generation for the next 24 hours [20]. The load demand increased significantly from 6:00 a.m. to 09:30 a.m. and was then maintained at a high level until 19:30 p.m. The WTs produced a large amount of power in the early morning and, in the case with no ESDs, the surplus power was exported via the substation bus. The PV arrays also contributed to the low-carbon-operation of the test grid during the daytime.

$$[\mathbf{P^k}, \mathbf{Q^k}]^T = \left[ \mathbf{O}_{1\times10}, \left[ \sum_s P_{i=1,s}^k, \sum_s Q_{i=1,s}^k \right], \mathbf{O}_{1\times8}, \left[ \sum_s P_{i=2,s}^k, \sum_s Q_{i=2,s}^k \right], \mathbf{O}_{1\times14}, \left[ \sum_s P_{i=3,s}^k, \sum_s Q_{i=3,s}^k \right], \ldots \right.$$

$$\left. \ldots \mathbf{O}_{1\times16}, \left[ \sum_s P_{i=4,s}^k, \sum_s Q_{i=4,s}^k \right], \mathbf{O}_{1\times4}, \left[ \sum_s P_{i=5,s}^k, \sum_s Q_{i=5,s}^k \right], \mathbf{O}_{1\times6} \right]^T, \quad \forall k \quad (27)$$

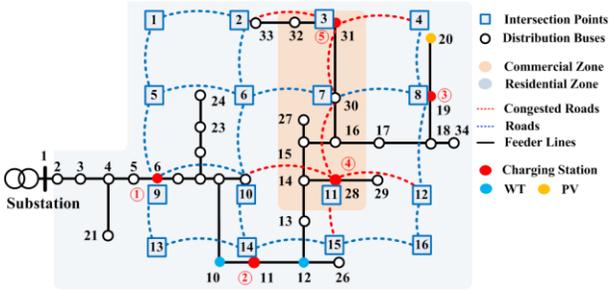

Fig. 4. Distribution and transportation networks for the case studies.

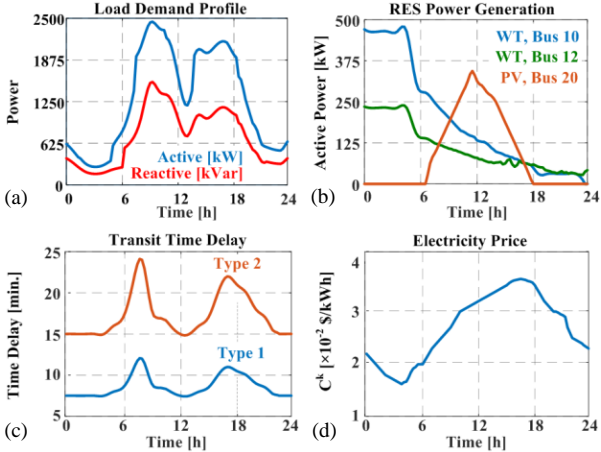

Fig. 5. (a) Total load demand in the distribution grid, (b) RES power generation, (c) transit time delay between adjacent intersections, and (d) electricity price.

TABLE I. MESD SPECIFICATIONS

| MESDs | $E_{s,min}$ [%] | $E_{s,max}$ [%] | $E_s^0$ [%] | $\Delta E_{s,max}$ [%] | $\eta_s$ [kWh/km] | $pf_{s,min}$ [pu] | $P_{s,max}$ [kW] | $P_{s,min}$ [kW] | $E_{s,cap}$ [MWh] |
|---|---|---|---|---|---|---|---|---|---|
| MESD$_1$ | 20 | 90 | 70 | 30 | 0.5 | 0.95 | 500 | −500 | 5.0 |
| MESD$_2$ | | | | | | | 188 | −188 | 2.5 |

TABLE II. CONGESTION TYPES FOR ROUTES BETWEEN INTERSECTIONS

| (a, b) | type | (a, b) | type | (a, b) | type | (a, b) | type | (a, b) | type |
|---|---|---|---|---|---|---|---|---|---|
| (2, 3) | 1 | (3, 7) | 2 | (7, 11) | 1 | (11, 10) | 2 | (12, 11) | 1 |
| (3, 2) | 2 | (4, 3) | 1 | (10, 11) | 1 | (11, 12) | 1 | (15, 11) | 1 |
| (3, 4) | 1 | (7, 3) | 1 | (11, 7) | 1 | (11, 15) | 1 | - | - |

TABLE III. FEATURES OF THE PROPOSED AND CONVENTIONAL STRATEGIES

| Strategies | | Devices | Mobility | Decision-making | Locations | $\Sigma_s E_{s,cap}/N_S$ |
|---|---|---|---|---|---|---|
| Prop. | Case 1 | MESDs | controllable | single-stage | co-optimized | large |
| Conv. | Case 2 | ESDs | stationary | two-stage | optimized in the 1st stage | large |
| | Case 3 | PEVs | driver dependent | two-stage | forecast in the 1st stage | small |

TABLE IV. PEV SPECIFICATIONS

| Number | $E_{s,min}$ [%] | $E_{s,max}$ [%] | $E_s^0$ [%] | $\Delta E_{s,max}$ [%] | $\eta_s$ [kWh/km] | $pf_{s,min}$ [pu] | $P_{s,max}$ [kW] | $P_{s,min}$ [kW] | $E_{s,cap}$ [MWh] |
|---|---|---|---|---|---|---|---|---|---|
| 22~38 | 20 | 90 | 40~80 | 30 | 0.25 | 0.95 | 18.1 | −18.1 | 0.197 |

Furthermore, Fig. 4 shows that the test grid contains five MESD charging stations, located on Buses 6, 11, 19, 28, and 31, respectively. In the proposed strategy, two MESDs travelled between stations to provide V2G services; the specifications on the MESDs are listed in Table I. Moreover, Fig. 4 shows the overlap with the transportation network, in which 16 intersections are distributed throughout residential and commercial zones. The stations are located at intersections 3, 8, 9, 11, and 14 on the road map. The distances between the adjacent intersections were all set to 1.5 km [15] for simplicity. We also considered two types of traffic congestion between two intersections, at least one of which includes intersection 3 or 11, as shown in Fig. 4. Table II lists the types of traffic congestion and Fig. 5(c) shows the forecast profiles of the corresponding transit time delays [13]. Note that routes that only passed through residential zones were not subject to traffic congestion. Fig. 5(d) shows the profile of the electricity price used to estimate $J$ in (19) [19]. The operating profiles, shown in Fig. 5, are commonly observed in various power and transportation networks, rather than specific types of networks; consequently, the case studies can be conducted without loss of generality.

The comparative case studies for the proposed and conventional strategies were conducted under various power and transportation network conditions: network size, traffic time delay, power and energy capacity, and incremental changes in the bus voltage and line power flow. Table III shows the main features of the proposed (i.e., Case 1) and conventional strategies (i.e., Cases 2 and 3). Because this paper focuses on the single-stage decision-making model and the controllable mobility, the conventional strategies have been defined as the two-stage model for stationary ESDs and PEVs.

Due to the different features, the objective functions and constraints in the conventional strategies were different from those in the proposed strategy. Specifically, in Case 2, the optimal locations and power outputs of the stationary ESDs were determined in the first and second stages, respectively [11], [13]. The optimal locations (i.e., $m_{is}^k$) were determined considering a 1-year profile of load demand [20], [21]. The optimal placement problem was formulated using the first term of (19) (i.e., $\Sigma_k C^k \cdot \Delta P_{loss}^k \cdot t_{unit}$) as the objective function and (20)–(30) as the constraints. The decision variables were set as $m_{is}^k$, while $e_{ijs}^k$ were fixed at zero. Then, given the optimal values of $m_{is}^k$, the optimization problem for the optimal power dispatch was formulated in the second stage where the objective function and constraints remained the same with those for the first-stage problem. The decision variables changed to $P_{is}^k$ and $Q_{is}^k$. In Case 3, the driver-dependent transit paths and times (or, equivalently, $m_{is}^k$ and $e_{ijs}^k$) of the individual PEVs were forecasted in the first stage based on historical data [22], [23]. Then, given the forecasted values of $m_{is}^k$ and $e_{ijs}^k$, the optimal power dispatch problem was formulated in the second stage, where the objective function, constraints, and decision variables were the same as those for Case 2.

The rated power and energy capacities of the stationary ESDs were set as the same as those of the MESDs. Table IV shows the PEV specifications [24]. For simplicity, the individual PEVs were assumed to have the same specifications. Due to the smaller battery size, the number of PEVs was set to a value ranging from 22 to 38 in the case studies, so that their total energy capacity was the same as those of the MESDs and the stationary ESDs.

The case study results were obtained by applying the CPLEX MILP solver to the optimization problems for the proposed and conventional strategies, which were formulated using the mathematical models of the IEEE test feeders.

## B. Optimal output power and transit path schedules of MESDs

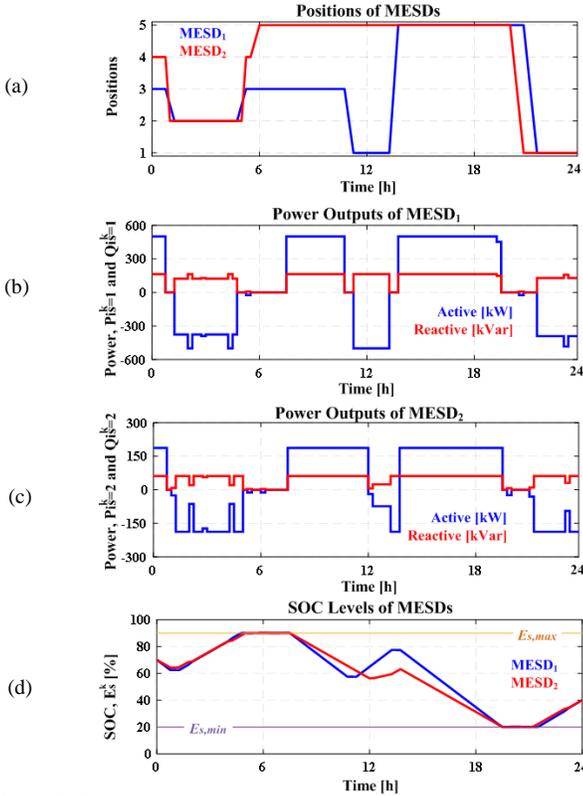

Fig. 6. Optimal schedules for the proposed strategy: (a) the stations where the MESDs are connected, (b), (c) MESD power outputs, and (d) SOC levels.

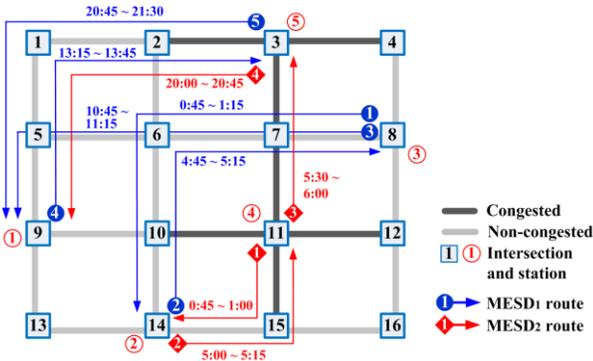

Fig. 7. Optimal routes and transit times of the MESDs for the proposed strategy.

TABLE V. COMPARISONS BETWEEN THE PROPOSED (CASE 1) AND CONVENTIONAL STRATEGIES (CASES 2 AND 3)

| Energy loss and cost | Proposed (1) Case 1 | Conventional (2) Case 2 | (3) Case 3 | (4) No ESD | Reduction rates [(2)–(1)]/ (2) | [(3)–(1)]/ (3) |
|---|---|---|---|---|---|---|
| $E_{loss,tot}$ [kWh] | 1932.0 | 2392.3 | 2560.4 | 3954.4 | 19.2% | 24.5% |
| $J$ [\$] | 59.1 | 70.3 | 77.7 | 124.1 | 15.9% | 23.9% |

Fig. 6(a) shows the optimal positions of the MESDs over 24 hours according to the proposed strategy: i.e., the stations to which the MESDs were connected. The corresponding transit routes are presented in Fig. 7. Furthermore, Figs. 6(b)–(d) show the optimal profiles of the power outputs and SOC levels of the MESDs. The MESDs initially provided a V2G service at stations 3 and 4, located far from the substation bus, to compensate for the relatively high active and reactive load demand until 0:45 a.m. The load demand decreased, while the power outputs of the WTs were maintained at high levels. The MESDs then moved to station 2, located between two WT buses, and absorbed a large amount of WT output power during the period from 1:00 a.m. to 4:45 a.m. Fig. 6(d) shows that the SOC levels increased to the maximum limit $E_{s,max} = 90\%$. This enabled the DSO to store renewable energy in MESD batteries and use it for power balancing during on-peak hours, significantly reducing network power loss.

The MESDs moved back to stations 3 and 5, close to the end buses of the main feeder, before the traffic delays started increasing rapidly at 6 a.m., as shown in Fig. 5(c). In other words, the MESDs changed locations beforehand, avoiding traffic congestion and preparing for increases in load demand and electricity price. The MESDs then supplied active and reactive power in V2G mode until 11 a.m. and 12 p.m., respectively, reducing the line power losses.

The load demand decreased over lunchtime. The MESDs then operated in charging mode to restore the SOC levels; the aim was to be ready for the provision of V2G services until 19:30 p.m. For charging, the MESD$_1$ moved from stations 3 to 1, close to the substation bus, to minimize the incremental increase in the power loss. For the transition, the MESD$_1$ took the fastest path, including intersections 8, 7, 6, 5, and 9, as shown in Fig. 7. This validates the hypothesis that integrating the linear transit model with the constraints on the distribution grid operation enabled the MESDs to travel between stations in a timely manner, according to the traffic and grid conditions.

Fig. 6 shows that the energy stored in the MESD batteries was mainly exploited at station 5, to reduce the costs of energy losses (i.e., $J$ in (19)) during the period from 13:45 p.m. to 19:30 p.m., when the electricity price was maintained at a high level. The MESDs then became mainly stationary due to traffic congestion. After the congestion was relieved approximately at 20:00 p.m., MESDs were moved to station 1 so that they could absorb active power to ensure stable, continuous MESD operation during the next day (i.e., (26)). Note that reactive power was still supplied to mitigate the incremental increase in the power loss due to the charging power.

Table V shows the comparisons between the total energy losses (i.e., $E_{loss,tot} = \sum_k P_{loss}^k \cdot t_{unit} + \sum_k \sum_j \sum_i \sum_s \eta_s \cdot d_{ij}^k \cdot e_{ijs}^k$) and corresponding costs $J$ for Cases 1–3, in addition to a baseline case where no ESDs were available. In the baseline case, the energy loss was calculated using general power flow equations. For Case 1, $E_{loss,tot}$ was 19.2% and 24.5% smaller than those for Cases 2 and 3, respectively. Moreover, $J$ was reduced by 15.9% and 23.9%, compared to Cases 2 and 3, respectively.

## C. Optimal scheduling of MESD operations under different specifications and network conditions

TABLE VI. SPECIFICATIONS OF SEVEN MESDs

| MESD | $E_{s,min}$ [%] | $E_{s,max}$ [%] | $E_s^0$ [%] | $\Delta E_{s,max}$ [%] | $\eta_s$ [kWh/km] | $pf_{s,min}$ [pu] | $P_{s,max}$ [kW] | $P_{s,min}$ [kW] | $E_{s,cap}$ [kWh] |
|---|---|---|---|---|---|---|---|---|---|
| MESD$_1$ |  |  |  |  |  |  | 188 | –188 | 750 |
| MESD$_2$ |  |  |  |  |  |  | 156 | –156 | 750 |
| MESD$_3$ |  |  |  |  |  |  | 125 | –125 | 750 |
| MESD$_4$ | 20 | 90 | 70 | 30 | 0.2 | 0.95 | 63 | –63 | 750 |
| MESD$_5$ |  |  |  |  |  |  | 63 | –63 | 500 |
| MESD$_6$ |  |  |  |  |  |  | 31 | –31 | 500 |
| MESD$_7$ |  |  |  |  |  |  | 31 | –31 | 250 |

Additional case studies were performed to further validate the feasibility and effectiveness of the proposed strategy. For example, the strategy was implemented using the test bed for a scenario featuring more MESDs. The operating conditions of the test bed remained the same, as shown in Fig. 5; the specifications of the MESDs are listed in Table VI. Fig. 8 shows the corresponding optimal schedules of MESD operations. For brevity, the output power profiles of MESD units 1, 5, and 7 are provided; those of the other MESDs were similarly represented. The overall operations of the MESDs in Fig. 8 are consistent with those in Fig. 6, although the numerical results are rather different. For the scenario, $E_{loss,tot}$ for Case 1 was reduced by 9.1% and 18.1%, compared to those for Cases 2 and 3, respectively (see Table III). In Case 1, $J$ was also 11.0% and 20.2% smaller than those in Cases 2 and 3, respectively.

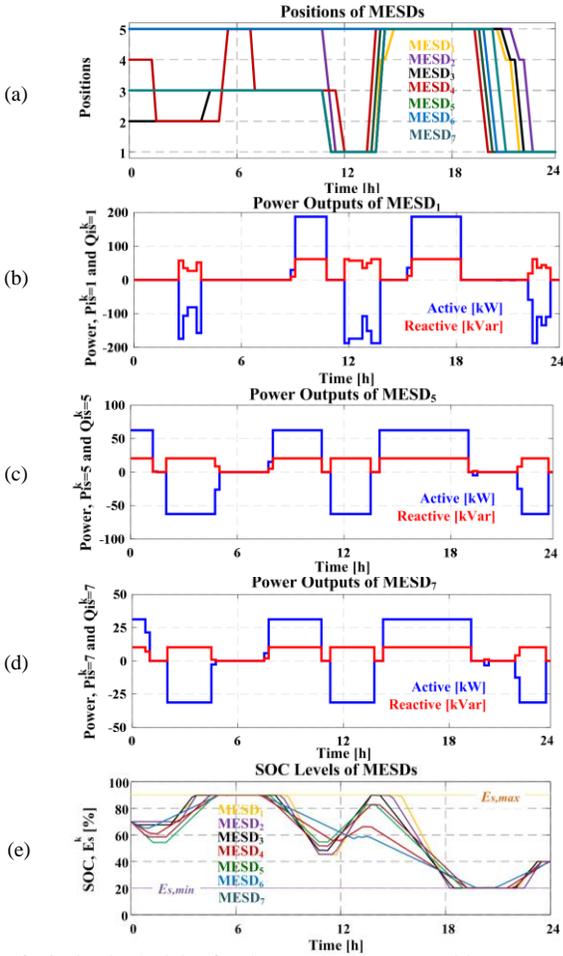

Fig. 8. Optimal schedules for the proposed strategy with $N_S = 7$: (a) MESD positions, (b)–(d) active and reactive power outputs, and (e) SOC levels.

TABLE VII. PROPOSED (CASE 1) AND CONVENTIONAL STRATEGIES (CASES 2 AND 3) FOR THE DIFFERENT ESD CONDITIONS, SHOWN IN TABLE VI

| Energy loss and cost | Proposed (1) Case 1 | Conventional (2) Case 2 | Conventional (3) Case 3 | Conventional (4) No ESD | Reduction rates [(2)–(1)]/(2) | Reduction rates [(3)–(1)]/(3) |
|---|---|---|---|---|---|---|
| $E_{loss,tot}$ [kWh] | 2509.5 | 2761.7 | 3062.6 | 3954.4 | 9.1% | 18.1% |
| $J$ [$] | 74.4 | 83.6 | 93.20 | 124.1 | 11.0% | 20.2% |

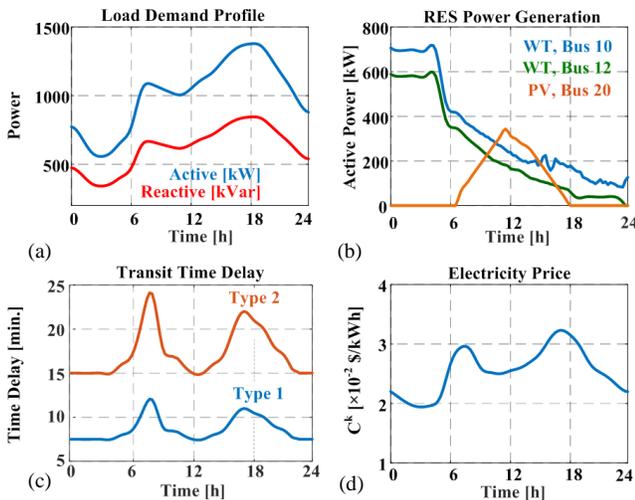

Fig. 9. A different set of network operating profiles: (a) total load demand in the distribution grid, (b) RES power generation, (c) transit time delay between adjacent intersections, and (d) electricity price.

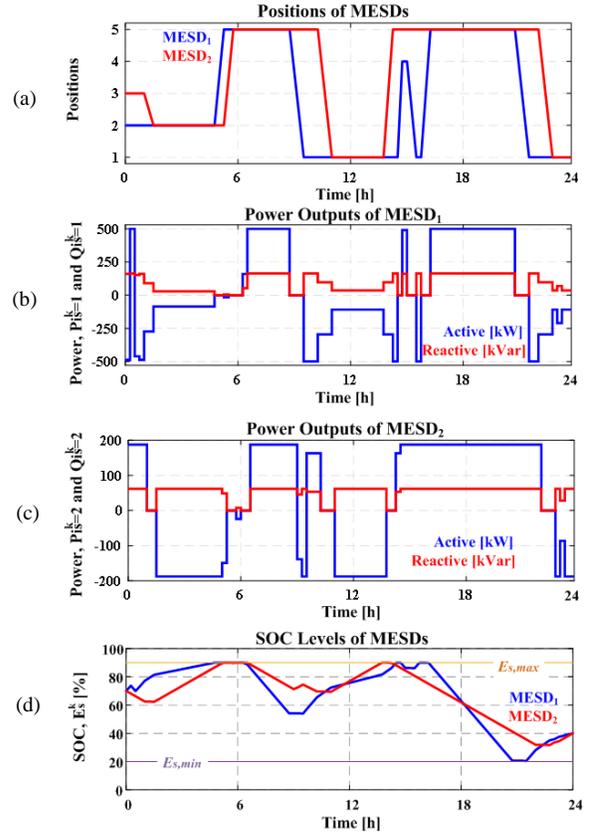

Fig. 10. Optimal schedules for the different network operating profiles, shown in Fig. 9: (a) MESD positions, (b), (c) active and reactive power outputs, and (d) SOC levels.

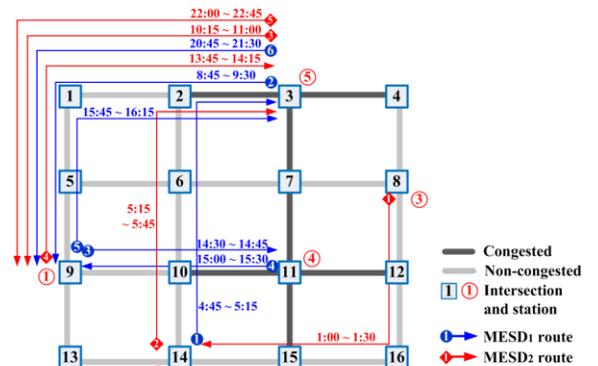

Fig. 11. Optimal routes and transit times of the MESDs for the different set of profiles shown in Fig. 9.

TABLE VIII. PROPOSED (CASE 1) AND CONVENTIONAL STRATEGIES (CASES 2 AND 3) FOR THE DIFFERENT NETWORK CONDITIONS, SHOWN IN FIG. 9

| Energy loss and cost | Proposed (1) Case 1 | Conventional | | | Reduction rates | |
|---|---|---|---|---|---|---|
| | | (2) Case 2 | (3) Case 3 | (4) No ESD | [(2)–(1)]/ (2) | [(3)–(1)]/ (3) |
| $E_{loss,tot}$ [kWh] | 1984.1 | 2589.8 | 2361.3 | 3493.0 | 23.4% | 16.0% |
| $J$ [$] | 49.8 | 64.1 | 62.28 | 94.9 | 22.3% | 20.0% |

Another case study was performed using the different network operating profiles, shown in Fig. 9, which were obtained from the data sources [13], [15], and [21], [25] for power and transportation networks in the same region (i.e., Los Angeles, CA, as an example of a large metropolitan area in the United States). The test bed configuration and the MESD specifications remained the same as those of Fig. 4 and Table I, respectively. Fig. 10 shows schedules for the optimal locations, power outputs, and SOC levels of the MESDs, which are similar to those in Fig. 6. The optimal routes and transit times of the MESDs in Fig. 11 are also similar to those in Fig. 7. Moreover, as shown in Table VIII, the proposed strategy still reduced $E_{loss,tot}$ and $J$, compared to the conventional strategies. In other words, the proposed strategy still enabled MESDs to travel between stations in a timely manner and minimized the total energy loss cost under time-varying conditions of the power and transportation networks. The consistent results of the case studies performed in Sections IV-B and IV-C confirm the effectiveness of the proposed strategy and validate its wide applicability under various conditions of the MESDs and the transportation and power networks.

### D. Effects of $P_{max}$, $E_{cap}$, $\Delta V_{max}$, and $\Delta L_{max}$ on optimal schedules

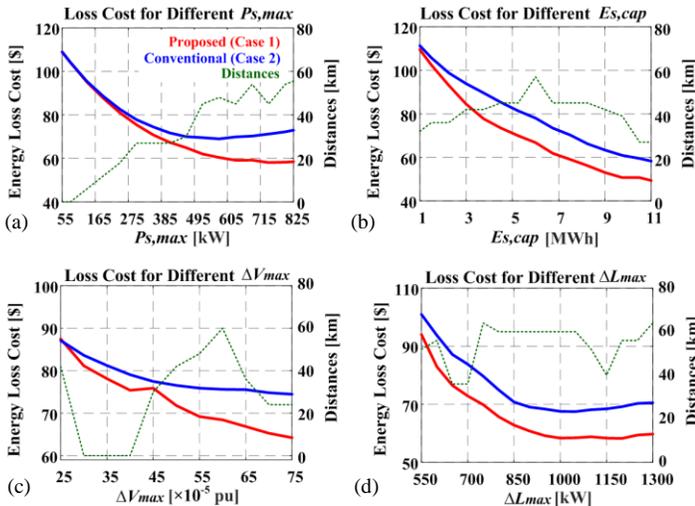

Fig. 12. The energy loss cost and distance moved with variation in, respectively, (a), (b) the MESD power and energy capacities and (c), (d) the maximum limits on incremental changes in bus voltage and line power flow.

The case study discussed in Section IV-B was repeated, as the rated power and energy capacities of the MESDs were gradually increased from $\sum_s P_{s,max}$ = 55 kW to 825 kW, and from $\sum_s E_{s,cap}$ = 750 kWh to 11 MWh, respectively. Fig. 12(a) and (b) show that the proposed strategy reduced the energy loss cost, $J$, in comparison to the conventional strategy for each value of $\sum_s P_{s,max}$ and $\sum_s E_{s,cap}$ tested. Note that the comparisons were made for Cases 1 and 2, where the rated power and energy capacities of the individual ESDs remained the same. Moreover, in the proposed strategy, $J$ was gradually reduced as $\sum_s P_{s,max}$ and $\sum_s E_{s,cap}$ increased, whereas in the conventional strategy, $J$ started to increase when $\sum_s P_{s,max}$ increased beyond approximately 577.5 kW because the power flow characteristics in the test grid changed significantly. The inflection value for the proposed strategy was expected to be larger than 577.5 kW, because the MESD mobility improved the flexibility of the grid operation. It also can be seen in Fig. 12(a) that, in general, the moving distance of the MESDs increased as $\sum_s P_{s,max}$ increased. This was mainly because the ratio of the decrease in the network energy loss to the increase in the moving energy loss increased. In other words, large savings in electric energy could be achieved for small transport energy costs.

Analogously, the case study discussed in Section IV-B was iterated for various maximum limits on the incremental changes in bus voltage and line power flow. For simplicity, all of the entries in $\Delta \mathbf{V_{max}^k}$ and $\Delta \mathbf{L_{max}^k}$ were assumed to be the same as $\Delta V_{max}$ and $\Delta L_{max}$, respectively. Fig. 12(c) and (d) show that the proposed strategy enabled the DSO to achieve more cost-effective, but still reliable, operation of the test grid than the conventional one for all values of $\Delta V_{max}$ and $\Delta L_{max}$. The difference in values of $J$ between the two methods increased gradually with $\Delta V_{max}$ and $\Delta L_{max}$: i.e., the constraints on the distribution grid operation became more relaxed. This implies that the DSO can obtain synergistic benefits by coordinating the MESDs, RESs, and voltage regulation devices, which can facilitate recovery of the investment costs for the purchase of MESDs. This possibility will be explored in future research.

### E. Application to large-scale distribution network

Fig. 13 shows the IEEE 123-Node Test Feeder, with the modifications based on [11], which was overlapped with a transportation network including three MESDs and eight charging stations. The results of the case study were similar to those described in Section IV-B. As shown in Fig. 14, the $MESD_{1, 2}$ charged the batteries in the morning to provide V2G service during the daytime. The charging was conducted at station 1 to mitigate the incremental increase in power loss. During the daytime, the MESDs then supplied both active and reactive power to compensate for the increase in load demand. As shown in Fig. 5(d), $C^k$ was also maintained at a high level. For better compensation, the MESDs moved to stations 6 and 7, near the end buses, at approximately 4:00 a.m. The MESDs took detours to station 8 to avoid delays due to traffic, which increased rapidly after 6 a.m. When the traffic congestion was relieved, the MESDs changed locations and operated in charging mode, restoring the SOC levels. The reactive power was still supplied to support the bus voltages and prevent large increases in power loss due to the charging power.

Fig. 14(c) shows that $MESD_3$ exhibited a rather different profile, complementing the other MESDs. For example, $MESD_3$ initially provided the V2G service at station 6, mitigating the influence of the charging power of the $MESD_{1, 2}$ on the line power loss. $MESD_3$ then moved to station 1 to charge its battery at 4:00 a.m., when $MESD_{1, 2}$ began V2G service provision. The MESD continued to travel between stations 1 and 6, because of the combined effects of the load demand variation and the complementary relationship between MESDs. The routes between the stations were not subject to traffic congestion, reducing the costs for the transits of MESD$_1$.

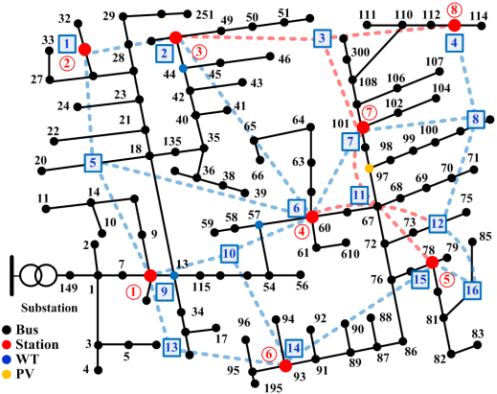

Fig. 13. Large-scale distribution grid used for the additional case studies.

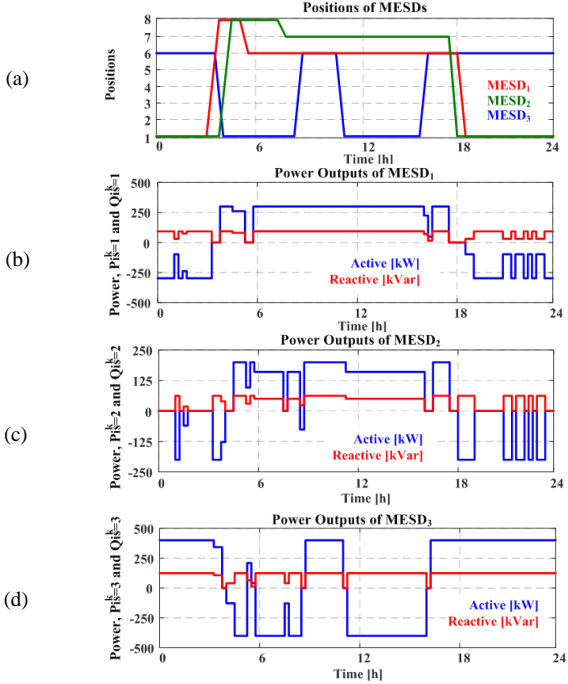

Fig. 14. Optimal schedules for the proposed strategy in the large-scale network: (a) the positions and (b)–(d) power outputs of the MESD$_{1, 2, 3}$.

TABLE IX. PROPOSED (CASE 1) AND CONVENTIONAL STRATEGIES (CASES 2 AND 3) FOR THE LARGE-SCALE NETWORK, SHOWN IN FIG. 13

| Loss, cost, and voltage | Proposed | Conventional | | | Reduction rates | |
|---|---|---|---|---|---|---|
| | (1) Case 1 | (2) Case 2 | (3) Case 3 | (4) No ESD | [(2)–(1)]/ (2) | [(3)–(1)]/ (3) |
| $E_{loss,tot}$ [MWh] | 12.7 | 17.9 | 19.2 | 23.6 | 29.1% | 33.9% |
| $J$ [$] | 390.3 | 549.3 | 596.1 | 723.3 | 28.9% | 34.5% |
| $V_{rms}$ [pu] | 0.9993 | 0.9990 | 0.9986 | 0.9981 | –0.03% | –0.07% |

Table IX lists the total energy losses and corresponding costs for the proposed and conventional strategies. It also shows the averages of the voltage magnitudes at the charging stations during the scheduling time period in the rms values: i.e., $V_{rms} = 1/N_I \cdot \sum_i (\sum_k (V_i^k)^2/N_K)^{1/2}$. In Case 2, three stationary ESDs were located at stations 6 and 8, and in Case 3, three PEVs were initially located at each charging station. The PEVs were assumed to depart from the initial locations and return to the locations via rather random transit paths. In the proposed strategy, $J$ was reduced by 28.9% and 34.5%, compared to the conventional strategies, due to the improved voltage support. Note that in Case 1, $V_{rms}$ increased by 0.03% and 0.07%, compared to Cases 2 and 3, respectively.

## V. DISCUSSION

The proposed scheduling strategy for V2G service provision is consistent with current practices in terms of ancillary service provision [26] over a scheduling time horizon (e.g., from minutes to hours or days) based on forecasts of system operating conditions. Various sophisticated models forecasting the operations of power and transportation networks one day ahead have been developed in previous studies (e.g., [27]–[30]); these can be readily integrated with the proposed strategy. Apart from such integration, a few changes are further required to enhance the robustness of the proposed strategy, particularly taking into consideration forecast errors and real-time uncertainty in power and transportation network operations. For example, short-term (i.e., from time-step-ahead to hour-ahead) forecast models of load demand, renewable power generation, electricity price, and transit time delays need to be developed for online estimation of differences between day-ahead and time-step-ahead forecast data, and consequently differences between day-ahead scheduling and the actual, real-time condition of MESD operations. Via online estimation, the DSO updates online the constants used in the optimization problem (1)–(30) and then re-schedules online the optimal power outputs and transit paths of the MESDs over a shorter scheduling period. The online estimation, parameter updating, and re-scheduling schemes are consistent with current practices in electricity market clearing [31], [32]. The integration of these schemes with the proposed strategy is a task for the future.

## VI. CONCLUSIONS

This paper proposed an optimal scheduling strategy where the power outputs and routes of MESDs are co-optimized to minimize the total cost incurred by energy losses in both the distribution grid and the transportation network. For optimal scheduling, an MESD transit model was developed using a set of linear equations with binary flags for the MESD connection, departure, and movement, taking into account the time-varying traffic congestion. The linear transit model could be readily integrated with the linearized constraints for stable grid operation. Under the proposed strategy, the optimization problem was formulated using linearized models and solved using an off-the-shelf MILP solver. The simulation case studies were carried out using IEEE test feeders, where the energy costs were reduced by 15.9% and 28.9%, respectively, in the proposed strategy, compared to the conventional one. In other words, the grid operational flexibility was effectively improved by the mobility of the MESDs: i.e., traveling around the stations adaptively according to time-varying load demands and traffic congestions. The case studies were repeated under a range of conditions, determined by varying factors such as the MESD power and energy capacities, and the maximum limit on the incremental changes in voltage magnitude and line power flow. The results of the case study confirmed that the proposed strategy successfully assists DSOs in taking advantage of MESDs to reduce energy loss costs, which can reduce the risks associated with investing in the MESD industry.